\documentclass[11pt]{article}
\usepackage{latexsym,bm,graphicx,color,xcolor,nicefrac,titletoc,enumerate,amsmath,amssymb,xfrac,xcolor,physics,cite,setspace}

\usepackage{mlmodern}
\usepackage[T1]{fontenc}

\usepackage[nottoc]{tocbibind} 

\usepackage{hyperref}
\hypersetup{linktocpage=true,colorlinks=true,linkcolor=blue,citecolor=blue,urlcolor=blue}

\usepackage[skip=3pt plus1pt, indent=20pt]{parskip}

\usepackage[bottom]{footmisc}

\usepackage{geometry}
\usepackage{titlesec}
\titleformat{\section}{\large\bfseries\sffamily}{\thesection}{0.5em}{}
\titleformat{\subsection}{\normalfont\bfseries\sffamily}{\thesubsection}{0.5em}{}
\titleformat{\subsubsection}{\normalsize\itshape\sffamily}{\thesubsubsection}{0.5em}{}
\titleformat*{\paragraph}{\normalsize\bfseries\sffamily}

\numberwithin{equation}{section}

\def\a{\alpha}

\def\f{\phi}
\def\vf{\varphi}

\def\k{\kappa}
\def\l{\lambda}

\def\m{\mu}
\def\n{\nu}

\def\nab{\nabla}
\def\pr{\prime}

\def\qq{\quad\quad}

\newcommand{\xp}{x_+}
\newcommand{\xin}{x_\text{in}}
\newcommand{\xout}{x_\text{out}}
\newcommand{\be}{\begin{eqnarray}}
	\newcommand{\ee}{\end{eqnarray}}
\newcommand{\lb}{\label}

\newcommand{\cH}{\mathcal{H}}

\newcommand{\cO}{\mathcal{O}}

\newcommand{\mail}[1]{\href{mailto:#1}{{\tt #1}}}

\begin{document}
	
		
		
	\begin{center}
		{\Large \bf \sffamily Simulating Hawking radiation in quantum many-body systems:\\ deviations from the thermal spectrum}
	\end{center}
		
	\begin{center}
		\vspace{10pt}
			
		{{\bf \sffamily G{\"o}khan Alka\c{c}}${}^{a}\,${\bf \sffamily  and Ege \"{O}zg\"{u}n}${}^{b}\,$}\\[4mm]
			
		{\small 
		{\it ${}^a$Department of Aerospace Engineering, Faculty of Engineering,\\ At{\i}l{\i}m University, 06836 Ankara, T\"{u}rkiye}\\[2mm]
		
		{\it ${}^b$Department of Physics Engineering,\\
			Hacettepe University, 06800, Ankara, T\"{u}rkiye}\\[2mm]
				
		{\it E-mail:} {\mail{alkac@mail.com}, \mail{egeozgun@hacettepe.edu.tr}}
		}
		\vspace{2mm}
		\end{center}
		
		\centerline{{\bf \sffamily Abstract}}
		\vspace*{1mm}
		\noindent We investigate a recently proposed one-to-one correspondence between quantum field theories in two-dimensional curved spacetime and quantum many-body systems, which enables the simulation of Hawking radiation in static background spacetimes. In particular, we demonstrate that deviations from the thermal spectrum, as predicted by the well-known tunneling method, can be observed in many-body simulations.
		\par\noindent\rule{\textwidth}{0.5pt}
		\tableofcontents
		\par\noindent\rule{\textwidth}{0.5pt}

\section{Introduction}
In the absence of a fully successful quantum theory of gravity, a reasonable approach is to employ a semi-classical approximation where we quantize matter fields on a curved background spacetime that is treated classically. Despite being merely an approximation, these quantum field theories (QFTs) in curved spacetimes have led to the prediction of quite a number of interesting phenomena such as Hawking radiation \cite{Hawking:1975vcx}, the Unruh effect \cite{Unruh:1976db,Davies:1974th} and particle creation in an expanding universe \cite{Parker:1968mv} (see \cite{Hollands:2014eia} for a review). 

On the other hand, since gravity is the weakest fundamental force, a direct observation of these phenomena is extremely difficult, if not impossible. However, Unruh's proposal to simulate Hawking radiation in a sonic analogue of a black hole \cite{Unruh:1980cg,Unruh:1994je} introduced a new avenue for exploration, and the study of analogue gravity models has become an important research program that has yielded valuable insights into the nature of quantum fields living in a curved spacetime (see \cite{Corley:1998rk,Barcelo:2000tg,Barcelo:2001ah,Schutzhold:2002rf,Basak:2002aw,Fedichev:2003id,Fedichev:2003bv,Barcelo:2003wu,Fischer:2004bf, Garay:1999sk, Weinfurtner:2010nu, Steinhauer:2015saa, MunozdeNova:2018fxv, Drori:2018ivu} for some seminal contributions in the field and \cite{Barcelo:2005fc} for a review).

In this work, we focus on a recently proposed framework \cite{Yang:2019kbb} that establishes a one-to-one correspondence between QFTs in 2d curved spacetime and quantum many-body systems. Starting from massless Klein-Gordon and Dirac equations in Eddington-Finkelstein coordinates, the authors find maps to three basic quantum many-body models: the hopping model, Hubbard model and XY model (see \cite{Tasaki:1995} for a review). An experimental realization of these maps in a system with trapped ions is also suggested. With this analogue gravity model, Hawking radiation and the entanglement properties of certain black hole spacetimes are successfully simulated.

As is well-known from the Unruh effect, the prediction that a uniformly accelerating observer in vacuum will experience a thermal bath, different observers observe different phenomena in curved spacetime. Therefore, it is of paramount importance to have a correspondence that admits the use of different coordinate systems. Also, one should always keep in mind that the ultimate aim is to simulate scenarios in 4d spacetime. Adressing these needs, the proposal of  \cite{Yang:2019kbb} was extended to different coordinate systems in 2d in \cite{Liu:2024wqj} and further generalized to a 3d case in \cite{Deger:2022qob}. 

In order to have a map to quantum many-body models, one needs to discretize the spatial coordinate of the evolution equations of matter fields. This discrete form is a reasonably good approximation for continuous fields when they vary slowly. Therefore, it is thought to be valid in the low-energy limit. Here, we will show that the framework of \cite{Yang:2019kbb} is actually capable of probing deviations from the thermal spectrum predicted by the tunneling method \cite{Parikh:1999mf}. It turns out that the choice of the metric function made in \cite{Yang:2019kbb} for convenience in numerical computations is an example with no corrections, and therefore, these corrections can only be probed with more ``realistic'' metric functions. The reader is invited to check \cite{Ribeiro:2021fpk,Ribeiro:2022gln} for recent investigations of the non-thermal features of Hawking radiation in the context of analogue gravity.

This paper is organized as follows: In Section \ref{sec:QFTtoQMB}, we review the essential elements of the proposal presented in \cite{Yang:2019kbb}. We emphasize that we do not claim any originality here; rather, we closely follow the logic and notation of \cite{Yang:2019kbb}. For the reader’s convenience, we repeat the key steps in the derivation of the Hamiltonian used in our simulations. In Section \ref{sec:Hawking}, we start with a detailed derivation of the emmision spectrum following \cite{Parikh:1999mf} and explain why the metric function used by the authors of \cite{Yang:2019kbb} is insensitive to corrections. Then, after a discussion of the corrected spectrum for the ``Schwarzschild'' black hole where the metric function is chosen as that of 4d Schwarzschild black hole, we give our numerical results for these two black hole backgrounds. We conclude our paper with a summary and outlook in Section \ref{sec:summary}.

\section{From massless Klein-Gordon equation to a bosonic hopping model}\label{sec:QFTtoQMB}
The Klein-Gordon equation for a massive complex scalar field $\f$ is given by
\begin{equation}\label{KG}
	(\nab^2+m^2) \f = 0,
\end{equation}
where $\nab^2 \equiv g^{\m\n} \nab_\m \nab_\n$ is the d'Alembert operator in curved spacetime, $g_{\m\n}$ is the metric and $\nab$ denotes its metric-compatible covariant derivative. As our background spacetime, we consider a 2d static spacetime whose line element can always be written in the Schwarzschild coordinates $(x,t)$ as
\begin{equation}\label{ds2}
	\dd{s}^2 = f(x) \dd{t}^2 - \frac{\dd{x}^2}{f(x)}.
\end{equation}
The event horizon is characterized by the zeros of the metric function, i.e. $f(\xp) = 0$. We assume that there is a single event horizon located at $x = \xp$. The Hawking temperature of the black hole is related to the surface gravity $\k = \frac{f^\pr(\xp)}{2}$ as $T = \frac{\k}{2\pi}$. Therefore, we have
\begin{equation}
	T = \frac{f^\pr(\xp)}{4 \pi}.
\end{equation}
Obviously, the line element \eqref{ds2} has a coordinate singularity at the horizon $x=\xp$, which can be removed by going to an in-falling Eddington-Finkelstein coordinates $(x,v)$ through
\begin{equation}
    \dd{t} \to \dd{v} - \frac{\dd{x}}{f(x)},
\end{equation}
such that the line element becomes
\begin{equation}
	\dd{s}^2 = f \dd{v}^2 - 2 \dd{v} \dd{x}.
\end{equation}
Now, the metric is well-defined at the horizon. In this coordinate system, the Klein-Gordon equation \eqref{KG} becomes
\begin{equation}
m^2 \phi-2 \partial_v \partial_x \phi-f^{\prime} \partial_x \phi-f \partial_x^2 \phi=0.
\end{equation}
By introducing new variables $\vf$ and $w$ as $m \varphi=2 \partial_v \phi+f \partial_x \phi$ and $w = \frac{\f}{\sqrt{f}}$, one obtains
\begin{equation}
\partial_v w=-\frac{f}{2} \partial_x w-\frac{f^{\prime}}{4} w+\frac{m \varphi}{2 \sqrt{f}}, \qq \partial_x \varphi=m w \sqrt{f} .
\end{equation}
In the massless case, one ends up with only one evolution equation given by
\begin{equation}\label{evol}
\partial_v w=-\frac{1}{4}\left[\partial_x(f w)+f \partial_x w\right] .
\end{equation}
It is possible to follow a similar procedure to show that the same evolution equation arises from the massless Dirac equation too. The crucial property of this equation is that since it is a first-order differential equation, it becomes possible to map it into the Heisenberg equation of a quantum many-body system after the quantization of the field $w$.

The next step is to discretize the system. For values of the spatial coordinate $x_n = n d$ with $n \in \mathbb{N}$ and $d \ll \l_0$, where $\l_0$ being the effective average wavelength in the system, the functions $f$ and $w$ are discretized as
\begin{equation}
f_n=f(x_n), \qq \qq w_n(v)=w\left(v, x_n\right).
\end{equation}
Using central differences to approximate $x$-derivatives and a variable transformation $w_n = (-i)^n e^{-i \mu v} \tilde{w}_n$, the evolution equation \eqref{evol} can be put into the following form
\begin{equation}
i \frac{\mathrm{~d}}{\mathrm{~d} v} \tilde{w}_n=-\kappa_n \tilde{w}_{n-1}-\kappa_{n+1} \tilde{w}_{n+1}-\mu \tilde{w}_n,
\end{equation}
where
\begin{equation}\label{kappa}
\kappa_n=\frac{f_n+f_{n-1}}{8 d} \approx \frac{f[(n-1 / 2) d]}{4 d}.
\end{equation}
At this point, $\m$ is an arbitrary constant. However, after quantization, it will play the role of a chemical potential and be set to zero to work in a setup where the particle number is conserved.

For quantization, the field $\tilde{w}_n$ is promoted to an operator. Since we consider a bosonic field, we take $\tilde{w}_n \to \frac{\hat{a}_n}{\sqrt{d}}$ and employ the usual bosonic commutation relations
\begin{equation}
\left[\hat{a}_n, \hat{a}_m^{\dagger}\right]=\delta_{n m}, \qq \left[\hat{a}_n, \hat{a}_m\right]=\left[\hat{a}_n^{\dagger}, \hat{a}_m^{\dagger}\right]=0.
\end{equation}
As a result, we obtain the following Heisenberg equation 
\begin{equation}
i \partial_v \hat{a}_n=\left[\hat{a}_n, \mathcal{H}\right],
\end{equation}
with the Hamiltonian
\begin{equation}\label{H}
\mathcal{H}=\sum_n\left[-\kappa_n\left(\hat{a}_n^{\dagger} \hat{a}_{n-1}+\hat{a}_{n-1}^{\dagger} \hat{a}_n\right)-\mu \hat{a}_n^{\dagger} \hat{a}_n\right],
\end{equation}
which describes a bosonic hopping model. Unlike real materials where the hopping parameter is constant, the effect of the curvature of the spacetime is reflected here by a site-dependent hopping parameter $\kappa_n$ given in \eqref{kappa}.

For the Dirac field, we promote the function into an anti-commuting operator and the resulting Hamiltonian is that of the free Hubbard model with again site-dependent hopping. By a Jordan-Wigner transformation, it is possible to rewrite it as an isotropic XY model with site-dependent hopping. The reader is referred to \cite{Yang:2019kbb} for details.

In order to simulate the black hole evaporation, we consider an initial state in the inner region of the black hole corresponding to a particle localized at the $n_0$-th site. We set $\mu=0$ since particle number is conserved. Then, we calculate the probability of finding a particle of energy $E$ in the outer region where $E$ is the positive eigenvalue of the Hamiltonian $\cH$ for the outer region. For the metric function $f(x) = \a \tanh x$ ($\a>0$: constant), the authors of \cite{Yang:2019kbb}  obtain an almost perfect match with the thermal spectrum
\begin{equation}
P \sim e^{-\frac{E}{T}},
\end{equation}
with the Hawking temperature $T=\frac{\a}{4 \pi}$ except at the low-energy region that is not expected to be covered due to the finite size cut-off. In the next section, we discuss the derivation of Hawking radiation by the tunneling method and present our numerical results, which demonstrate that the corrections to the thermal spectrum can be realized in many-body simulations.

\section{Hawking radiation as tunneling and numerical results}\label{sec:Hawking}
In standard derivations of Hawking radiation \cite{Hawking:1975vcx,Gibbons:1976ue}, one either studies the behavior of quantum fields by imposing appropriate boundary conditions or, treating the black hole immersed in a thermal bath, shows that a metastable equilibrium implying emission at Hawking temperature is possible. In these approaches, a fixed background geometry is considered.

On the other hand, it is quite obvious that a more realistic derivation requires a dynamical geometry admitting one to take the decrease in the black hole mass as it radiates into account. Such a derivation was given in \cite{Parikh:1999mf}, which also has the advantage that it is in accordance with the heuristic picture of viewing the source of radiation as tunneling.

In the tunneling method, the Boltzmann factor is related to the imaginary part of the action $S$ for the classically forbidden process as
\begin{equation}\label{PtoImS}
    P \sim e^{- 2 \Im[S]}.
\end{equation}
where the imaginary part of the action reads
\begin{equation}\lb{action}
    \Im[S] = \Im \left \{ \int_{x_{\text{in}}}^{x_{\text{out}}} p \dd{x} \right \}.
\end{equation}
In order to proceed, one again needs to get rid of the coordinate singularity in the line element \eqref{ds2} at the horizon $x = \xp$. This time, we switch to Gullstrand-Painlev\'e coordinates whose usefulness in the quantum mechanics of black holes was first emphasized in \cite{Kraus:1994fh}. After the following coordinate transformation
\begin{equation}
\dd{t} \rightarrow \dd{t}-\frac{\sqrt{1-f(x)}}{f(x)} \dd{x},
\end{equation}
the line element \eqref{ds2} becomes
\begin{equation}
\dd{s}^2=f(x) \dd{t}^2-\dd{x}^2-2 \sqrt{1-f(x)} \dd{x} \dd{t},
\end{equation}
 whose null geodesics are given by
 \begin{equation}
\dot{x}=\dv{f}{x}= \pm 1-\sqrt{1-f(x)},
\end{equation}
where $+$ and $-$ are valid for  outgoing and ingoing geodesics, respectively. This geodesic equation is modified by particle's self-gravitation. In \cite{Kraus:1994by}, it was shown that a shell of energy $E$ moves along geodesics with the replacement $M \to M+E$ when the black hole mass $M$ is kept constant and the ADM mass is allowed to vary. In our scenario, we want to conserve the total mass while the black hole mass decreases due to the emission of particles. Therefore, we should make the replacement $M \to M-E$ in describing the geodesics that will be used in our tunneling calculation.

The pivotal observation of \cite{Parikh:1999mf} is that the imaginary part of the action in \eqref{action} can be calculated without going into the details of the solution. The final expression for the imaginary part of the action that is valid for any metric function $f(x)$ is as follows
\begin{equation}\label{ImS}
\Im[S]=-\Im\left\{\int_0^E \dd{E^\pr} \int_{x_\text{in}}^{x_\text{out}} \frac{\dd{x}}{1-\sqrt{1-f(x)}\eval_{M \to M-E^\pr}} \right\}.
\end{equation}
The $x$ integral can be evaluated by analytical continuation to the complex plane. Using a counter-clockwise contour around the pole at $\xout$ yields the correct result if $\xin = \xp$ and $\xout = \xp\eval_{M \to M-E}$. This might seem a classically allowed path since $\xin>\xout$. On the contrary, it is forbidden since the horizon is contracting. The particle starts its motion from just inside the horizon and ends just outside the horizon.

We can now turn our attention to specific examples. In \cite{Yang:2019kbb}, the metric function is chosen as $f(x) = \alpha \tanh(x-\xp)$ and the location of the event horizon is set to zero ($\xp = 0$) for numerical convenience. For the time being, let us consider a nonzero $\xp$ and take it as $\xp = C\,M$ ($C>0$: constant), which is a reasonable form because, when $x$ is a ``radial'' coordinate ($x>0$), it describes the vanishing of the event horizon as $M \to 0$. Using \eqref{ImS}, we write the imaginary part of the action as
\begin{equation}
    \Im[S]=-\Im\left\{\int_0^E \dd{E^\pr} \int_{C M}^{C (M-E^{\pr})} \frac{\dd{x}}{1-\sqrt{1-\alpha \tanh[x-C(M-E^\pr)]}} \right\}.
\end{equation}
After evaluating the $x$ integral by residue theorem, we find
\begin{equation}
\Im[S]=\Im\left\{\frac{2 \pi i}{\alpha} \int_0^E \dd{E^{\prime}}\right\} = \frac{2 \pi E}{\alpha}.
\end{equation}
From \eqref{PtoImS}, we obtain
\begin{equation}\label{spec1}
\ln P \sim - \frac{E}{T}, \qq \qq T=\frac{\a}{4 \pi}.
\end{equation}
As we see, this is a very special case in the sense that although we have taken a dynamical background and imposed conservation of mass, the spectrum is still thermal and no corrections appear. In \cite{Yang:2019kbb}, this thermal spectrum was successfully reproduced in the many-body simulation. On the other hand, this choice for the metric function is not entirely satisfying since the temperature is independent of the black hole mass $M$ and does not yield the flat spacetime as $M \to 0$. Instead, one still has a black hole spacetime even with $M=0$ and $\xp =0$.

It would be interesting to study a more physical scenario. For this purpose, we choose the metric function as $f(x) = 1-\frac{\xp}{x}$ with $\xp = 2M$. Now, we recover the flat spacetime as $M \to 0$ and the temperature is mass dependent. This time, the imaginary part of the action is given by
\begin{equation}
    \Im[S]=-\Im\left\{\int_0^E \dd{E^\pr} \int_{2M}^{2(M-E^{\pr})} \frac{\dd{x}}{1-\sqrt{\frac{2(M-E^\pr)}{x}}} \right\}.
\end{equation}
From the residue theorem, we find
\begin{equation}
    \Im[S] = \Im \left\{4 \pi i \int_0^E \dd{E^\pr} (M - E^\pr)\right\}= 4 \pi E\left(M - \frac{E}{2}\right).
\end{equation}
Using \eqref{PtoImS}, we obtain
\begin{equation}\label{spec2}
\ln P \sim - \frac{E}{T} + 4 \pi E^2, \qq \qq T=\frac{1}{8 \pi M}.
\end{equation}
This is just the calculation of \cite{Parikh:1999mf} in 2d. Since we have taken the metric function as that of 4d Schwarzschild black hole, we arrive at the same corrected spectrum. 

\begin{figure}[h!]
\centering
\includegraphics[scale=0.50]{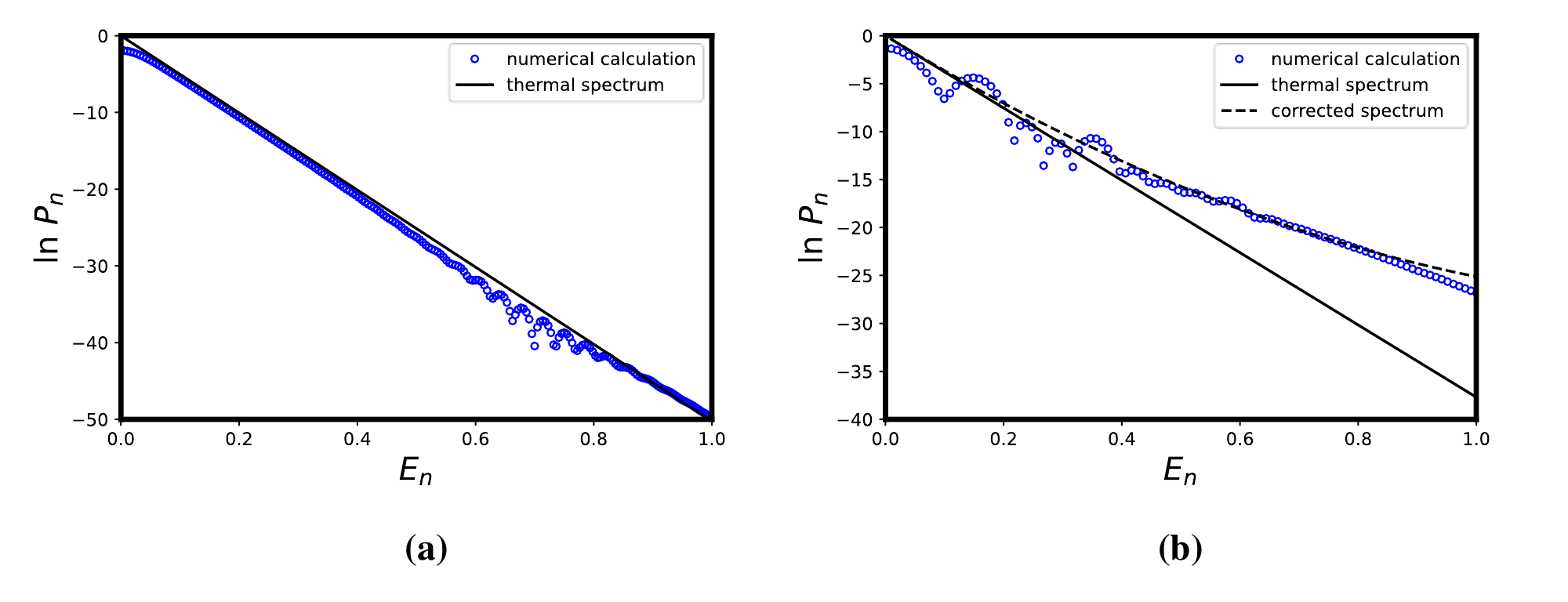}
\caption{Probability of finding a particle with energy $E_n$ outside the black hole calculated from the Hamiltonian $\cH$ given in \eqref{H}. {\bf a)} Result for $f(x)=\alpha \tanh(x-x_+)$ reproducing the results of \cite{Yang:2019kbb}. As explained in the main text, this choice for the metric function is a special case where the thermal spectrum remains intact.  {\bf b)} Result for $f(x)=1-x_{+}/x$ where the correction from the tunneling method is clearly visible. Blue circles represent the results of the numerical calculations. Full lines denote the thermal spectrum with the corresponding Hawking temperatures and the dashed line is the corrected spectrum.}
\label{figs}
\end{figure}

The moral of the story is that the choice of \cite{Yang:2019kbb} is insensitive to corrections that normally appear when working with a physical metric. A natural question is whether these corrections can also be realized in many-body simulations. The answer is affirmative. Our numerical results obtained by the simulation described in Section \ref{sec:QFTtoQMB} are presented in Figure \ref{figs}.

In order to make a comparison between the metric functions $f(x) = \alpha \tanh(x-\xp)$ and $f(x)=1-\frac{\xp}{x}$, we use the same range for the energy $E$ of the emitted particles as $0\leq E \leq 1.0$. For $f(x) = \alpha \tanh(x-\xp)$, we obtain the thermal spectrum \eqref{spec1} with acceptable numerical errors. On the other hand, for $f(x)=1-\frac{\xp}{x}$, we obtain the corrected spectrum in \eqref{spec2}, again with reasonable numerical accuracy, instead of a thermal spectrum. Note that this metric function was shown to exhibit a thermal spectrum in \cite{Liu:2024wqj}. However, the energy range for the emitted particles in that simulation is $0 \leq E \leq 0.03$, where the $E^2$-correction in the spectrum is not visible. Here, thanks to a wider energy range, we are able to show that the many-body simulations are able to capture the deviations from the thermal spectrum predicted by the tunneling method of \cite{Parikh:1999mf}. 

\begin{figure}[h!]
\centering
\includegraphics[scale=0.40]{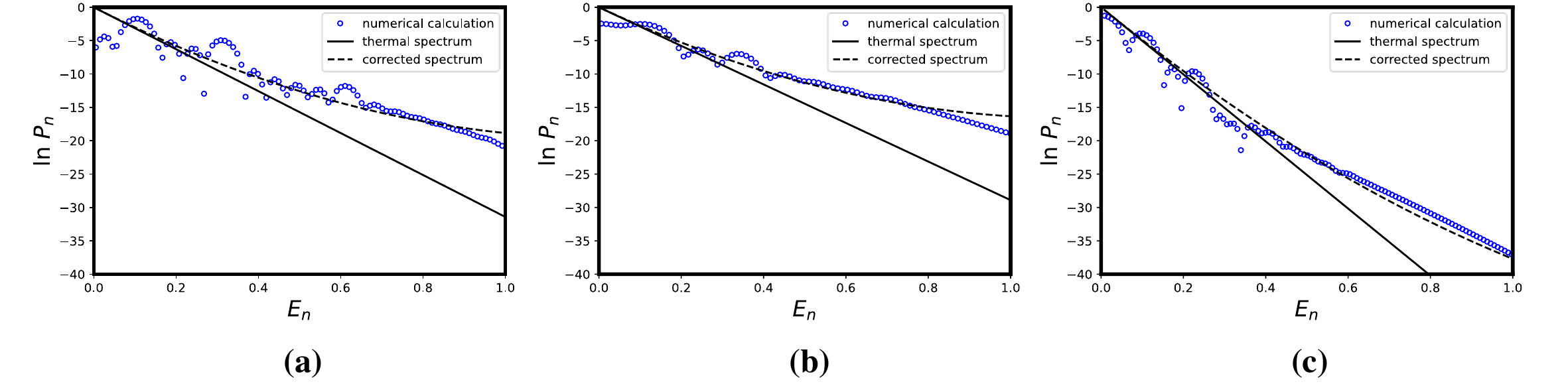}
\caption{Numerical results obtained with different parameters given in Table \ref{tab:parameters} for $f(x)=1-\frac{\xp}{x}$.}
\label{figs2}
\end{figure}

To demonstrate the validity of our numerical calculations for $f(x)=1-\frac{\xp}{x}$, different parameter ranges are covered as given in Figure \ref{figs2}. Parameters used in the numerical simulations are given in Table \ref{tab:parameters}. $L$ sets the number of sites such that the total number of sites are given by $N_s=2L+1$. $x_+$ is the location of event horizon, as explained before, and $n_0$ is the initial position of the particle. While $d$ is the lattice spacing, $t_{\text{max}}$ and $\delta t$ are the the final time value for the simulation and the time step size in the numerical solution of the Schr\"odinger equation respectively. The parameter $\alpha$ is only relevant for the case $f(x) = \alpha \tanh(x-\xp)$. The upper bound for energy eigenvalues for the case $f(x) = 1-\frac{\xp}{x}$ is given by $E_n \ll \cO(\nicefrac{1}{d})$. Thus, one expects the simulation results to be valid upto $E_n \sim \cO(1)$ for $d=0.1$, which is consistent with our numerical results. The small deviations at low energies are due to finite size of the lattice used in numerical calculations as also pointed out in \cite{Yang:2019kbb}. Finally, the relevant time values to capture the correct dynamics depend on the number of sites and lattice spacing as: $t_{\text{max}} \leq \cO(N_s d)$.

It is also worthy to mention the small numerical errors close to energies $E_n \sim \cO(1)$ from the thermal spectrum in our calculations for the case $f(x) = \alpha \tanh(x-\xp)$ shown in Figure \ref{figs}a. The effect of $\alpha$ to the upper bound of energy is given by $E_n \ll \cO(\nicefrac{\alpha}{d})$. Contrary to the parameter choices of $\alpha=10$ and $d=0.1$ in \cite{Yang:2019kbb}, which corresponds to the upper bound $E_n \sim \cO(100)$,  for our calculations we choose $\alpha=0.25$. The reason behind this choice is to capture as many eigenvalues as possible in the energy interval $E \in [0,1]$ to make a consistent comparison with the case $f(x)=1-\frac{\xp}{x}$ given in Figure \ref{figs}b.  Thus, with our choice of parameters, some numerical errors are expected.    

\setcounter{figure}{0}
\begin{table}[h!]
\caption{Parameters used in numerical simulations}
\centering
\vspace{0.2cm}
\begin{tabular}{||c c c c c c c c||} 
 \hline
 Figure & $L$ & $x_+$ & $n_0$ & $d$ & $t_{\text{max}}$ & $\delta t$ & $\alpha$ \\ [0.5ex] 
 \hline\hline
 1a & 700 & 0 & -20 & 0.1 & 200 & 0.1 & 0.25 \\ 
 1b & 700 & 30 & 10 & 0.1 & 120 & 0.1 & - \\
 2a & 700 & 25 & 5 & 0.1 & 115 & 0.1 & - \\
 2b & 600 & 23 & 0 & 0.1 & 80 & 0.1 & - \\
 2c & 800 & 40 & 20 & 0.1 & 150 & 0.1 &- \\ [1ex] 
 \hline
 \end{tabular}
\label{tab:parameters}
\end{table}

\section{Summary}\label{sec:summary}
In this paper, we have studied the proposal of \cite{Yang:2019kbb} for the simulation of Hawking radiation in static 2d spacetimes using quantum many-body systems. We have shown that while the choice made by the authors for the metric function is enough to demonstrate the success of the proposed map, it is a special case where there is no deviation from the thermal spectrum. By making a more physical choice in the sense that there is a deviation, as expected from conservation of energy during the emission, we have shown that the corrected spectrum is visible in many-body simulations.

By the tunneling method of \cite{Parikh:1999mf}, it is possible to derive the emission spectrum for any static black hole spacetime. Therefore, an obvious next step is to test the validity of the map of \cite{Yang:2019kbb} for different choices of the metric function, including those that admit multiple horizons. Also, in 4d, the imaginary part of the action for the classically forbidden process is related to the Bekenstein-Hawking entropy as $\Delta S_{\text{BH}} = -2 \Im[S]$. Therefore, it might be interesting to realize the black hole spacetimes studied here as exact solutions of 2d gravities to check whether this relation still holds.
\paragraph*{Acknowledgements} E. \"{O}. is grateful for discussions with Run-Qiu Yang on the details of the numerical calculations. E. \"{O}. acknowledges funding from the Scientific and Technological Research Council of Turkey (T\"{U}B\.{I}TAK) under Project Number 122F336. G. A. is supported by T\"{U}B\.{I}TAK under Project Number 124F058.

\end{document}